\documentclass[prc,twocolumn,showpacs,showkeys,nofootinbib,preprintnumbers,amsmath,amssymb]{revtex4}

\usepackage{graphicx}
\usepackage{dcolumn}
\usepackage{bm}

\begin{document}

\title{ Anisotropic flow of charged and identified hadrons in the 
quark-gluon string model\\
for Au+Au collisions at $\bf \sqrt{s_{NN}} = 200$ GeV }

\author{G. Burau$^1$}
\author{J. Bleibel$^1$}
\author{C. Fuchs$^1$}
\author{Amand Faessler$^1$}
\author{L.V. Bravina$^{2,3}$}
\author{E.E. Zabrodin$^{2,3}$}

\affiliation{
$^1$ Institute for Theoretical Physics, University of T\"ubingen, 
Auf der Morgenstelle 14, D-72076 T\"ubingen, Germany\\
$^2$ Department of Physics, University of Oslo, 
Blindern 1048, N-0316 Oslo, Norway\\
$^3$ Skobeltzyn Institute for Nuclear Physics, Moscow State University, 
Vorobievy Gory, RU-119899 Moscow, Russia
}

\date{February 22, 2005}

\begin{abstract}
The pseudorapidity behaviour of the azimuthal anisotropy parameters 
$v_1$ and $v_2$ of inclusive charged ($h^{\pm}$) hadrons and their 
dependence on the centrality has been studied in Au+Au collisions 
at full RHIC energy of $\sqrt{s_{NN}} = 200~{\rm GeV}$ within the 
microscopic quark-gluon string model. 
The QGSM simulation results for the directed flow $v_1$ show antiflow
alignment within the pseudorapidity range $|\eta| \leq 2$ in 
a fair agreement with the experimental $v_1(\eta)$ data, 
but cannot reproduce the further development of the antiflow up to
$|\eta| \approx 3.5$. 
The $\eta$ dependence of the elliptic flow $v_2$ extracted from 
the simulations agrees well with the experimental data in the
whole pseudorapidity range for different centrality classes.
The centrality dependence of the integrated elliptic flow of charged
hadrons in the QGSM almost coincides with the PHOBOS experimental 
distribution.
The transverse momentum dependence of the elliptic flow of identified
($\pi^{\pm}, K^{\pm}, p, \bar{p}$) and inclusive charged hadrons is
studied also. The model reproduces quantitatively the 
low-$p_T$ part of the distributions rather good, but underestimates
the measured elliptic flow for transverse momenta $p_T > 1$ GeV/$c$.
Qualitatively, however, the model is able to reproduce the saturation
of the $v_2(p_T)$ spectra with rising $p_T$ as well as the crossing 
of the elliptic flow for mesons and baryons.
\end{abstract}



\maketitle


\section{Introduction}
\label{intro}

Ultra-relativistic heavy ion collisions (URHICs) offer a unique
opportunity to study the nuclear phase diagram at high temperatures
and densities \cite{QM04}. The matter under such extreme conditions
has probably existed in the early universe within the first few
fm/$c$'s after the Big Bang. Therefore, it is very tempting to
investigate the properties of the {\it Little Big Bang\/}
\cite{NucBigBang} in the laboratory, and to search for a
new state of matter, predicted by the fundamental theory of strong
interactions (Quantum Chromodynamics -- QCD), namely, a plasma of
deconfined quarks and gluons (QGP). 

Among the various experimental studies of URHICs operates the 
Relativistic Heavy Ion Collider (RHIC) at BNL since 2000 to 
investigate gold on gold collisions up to $\sqrt{s_{NN}} = 200$ GeV.
After four years of operation strong experimental evidence has been
accumulated that at RHIC energies indeed a new state of matter is
created which is qualitatively different from a hadron gas
(see \cite{QM04} and references therein). This state seems, however,
not to behave like a weakly interacting parton gas - as could have
been naively expected - but rather like a strongly coupled plasma.
One argument towards such a scenario is the large elliptic flow
observed at RHIC \cite{ellST01,PHOB_fl,BackPHOBOS:2004,Adler2003}.
The development of strong elliptic flow requires short equilibration
times and large pressure gradients to drive the dynamics, both being
characteristic features of a strongly interacting system.

An independent argument for such a scenario is provided by lattice
QCD, though still on a qualitative basis. While lattice QCD predicts
undoubtedly a QGP phase transition around a critical temperature of
$T_c\sim 150\div 180~{\rm MeV}$ the properties of such a state are not
yet so well understood. As a striking fact lattice calculations
do not reach the Stefan-Boltzmann limit for the pressure of a free
parton gas $p\propto T^4$ even at $T \geq 5 T_c$ but saturate around 
80\% of the Stefan-Boltzmann limit \cite{karsch02}. Taking this fact 
seriously and not as a lattice artefact, one may argue that the QGP 
at temperatures well above the critical one is still a strongly 
interacting system \cite{shuryak04}.

In line with these arguments goes the success of hydrodynamics in
describing the collective flow at RHIC. Only at incident energies of
$\sqrt{s_{NN}} \geq 130~{\rm GeV}$ (which corresponds to an initial 
energy deposit of about $10~{\rm GeV/fm^3}$ in the central reaction 
zone) the hydrodynamic limit for the elliptic flow seems to be 
reached \cite{voloshin03}. Within the limits of tuning of the equation 
of state (including incorporation of a QGP phase transition) 
hydrodynamic calculations are able to describe the bulk properties of 
the collective flow rather reasonable \cite{Kolb:2000fh,HKHRV2001,Ret04}.
However, when turning to more differential observables such as 
centrality dependence, mass dependence or $p_T$ dependence of the 
elliptic flow, hydrodynamic calculations have also some problems 
to match the data.

In this context it is also important to obtain an understanding of the
reaction dynamics in terms of microscopic models. One class of
microscopic models which have been very successfully applied at CERN
SPS energies and below are string models. 
The main assumption of string models is that hadrons are produced as a
result of excitation and decay of open strings with different quarks 
or diquarks at their ends. 
Generally, all models are formulated as Monte Carlo event 
generators allowing to perform a careful analysis of the measurable
quantities by introducing all necessary experimental cuts. 
There are numerous versions of the two basic string-motivated
phenomenological approaches: the FRITIOF model \cite{Fritiof} and
the Dual Parton Model (DPM) \cite{dpm}. These two approaches
use different mechanisms of string excitation. In the first approach, 
relied on relativistic classical string dynamics, the string
masses arise from momentum transfer. The second one is based on the
Gribov-Regge theory (GRT) \cite{GRT} in the framework of relativistic 
quantum theory where quantum aspects like unitarity play an essential 
role. Here, the strings are produced as a result of colour exchange.

By construction, such types of models do not contain explicitly a 
quark-hadron phase transition. However, during the temporal evolution 
of a heavy ion reaction a dense and strongly interacting plasma is 
formed within such approaches as well. The system consists of partons 
and colour-flux tubes (or strings).
Thus, it is an essential question if string models are
able to create a sufficient amount of pressure in order to produce
large elliptic flow seen at RHIC, which features
they can describe and where they might fail. Such investigations
are in particular relevant since transverse as well as elliptic 
flow at SPS energies is well reproduced within the string-cascade
approach \cite{DF_prc00,DF_prc01,LPX99,soff99}.
By such type of studies one can
obtain deeper insight in the question which observables indicate
the appearance of {\it new} physics not included in standard
approaches to relativistic heavy ion collisions.

In the present work we describe ultra-relativistic heavy ion 
reactions by a microscopic quark-gluon string cascade model (QGSM) 
based on Gribov-Regge theory. Details of the model are given in the
next section. The QGSM has been demonstrated \cite{ell_flow} 
to give a fair description of the first data of 
$v_2$ in Au+Au reactions at $\sqrt{s_{NN}}=130$ GeV \cite{ellST01}. 
Here, we extend the analysis to full RHIC energy 
($\sqrt{s_{NN}}=200$ GeV) 
and to more differential observables, i.e. centrality dependence, 
mass dependence and $p_T$ dependence of $v_1$ and $v_2$.


\section{Quark-gluon string model}
\label{mcqgsm}

The Quark-Gluon String model (QGSM) incorporates partonic and
hadronic degrees of freedom and is based on GRT accomplished by a
string phenomenology of particle production in inelastic
hadron-hadron ({\it hh\/}) collisions. To describe hadron-hadron,
hadron-nucleus and nucleus-nucleus collisions the cascade procedure
of multiple secondary interactions of hadrons was implemented. The
QGSM incorporates the string fragmentation, formation of resonances,
and rescattering of hadrons, but simplifies the nuclear effects
neglecting, e.g., the mean fields or evaporation from spectators.
As independent degrees of freedom the QGSM includes octet and decuplet
baryons, octet and nonet vector and pseudoscalar mesons, and their
antiparticles. The momenta and positions of nucleons inside the
nuclei are generated in accordance with the Fermi momentum
distribution and the Woods-Saxon density distribution, respectively.
Pauli blocking of occupied final states is taken into account.
Strings in the QGSM can be produced as a result of the colour
exchange mechanism or, like in diffractive scattering, due to
momentum transfer. The Pomeron, which is a pole with an intercept
$\alpha_P(0) > 1$ in the GRT, corresponds to the cylinder-type
diagrams. The $s$-channel discontinuities of the diagrams, 
representing the exchange by $n$-Pomerons, are related to the process
of $2 k\, (k\leq n)$ string production. If the contributions
of all $n$-Pomeron exchanges to the forward elastic scattering
amplitude are known, the Abramovskii-Gribov-Kancheli (AGK) cutting
rules \cite{agk} enable one to determine the cross sections for
$2k$-strings. Hard gluon-gluon scattering and semi-hard processes
with quark and gluon interactions are also incorporated in the model
\cite{hard}. The inclusive spectra in the QGSM have automatically the
correct triple-Regge limit for the Feynman variable $x \rightarrow 1$,
double-Regge limit for $x \rightarrow 0$, and satisfy all conservation 
laws \cite{qgsm1}.
The particular stages of the collision model, namely 
(i) initialization of interacting projectile and target nuclei, 
(ii) string formation via inelastic nucleon-nucleon (hadron-hadron)
interaction, (iii) string fragmentation, i.e. hadronization, and 
(iv) hadron-hadron rescattering, are solved basically by Monte Carlo 
simulation techniques \cite{qgsm2}.


\section{Anisotropic flow of inclusive charged and identified hadrons}
\label{flow}

Among the main signals, which can help to reveal the formation of the
QGP in the experiment, are collective flow phenomena. Flow is directly 
linked to the equation of state of the excited matter. Generally, an 
effective EOS extracted from the model studies shows ``softness'' 
during the early stages of the collision, but it remains unclear 
whether the observed softness is (i) due to the proximity of the QCD 
phase transition \cite{HuSh95,RiGy96,Br94,CsRo99}, or (ii) due to 
non-equilibrium phenomena, such as the formation and fragmentation of 
strings \cite{Am91,Sorprl97}, or (iii) due to nuclear shadowing 
\cite{DF_prc00,DF_prc01,Br95}.

The transverse collective flow can be subdivided into isotropic 
and anisotropic flow. Two types of anisotropic flow, that are 
characterized by the first and the second harmonic coefficients 
of the Fourier decomposed invariant azimuthal distribution 
in momentum space \cite{VoZh96,PoVo98}
\begin{equation}
E \frac{d^3N}{d^3p} = \frac{1}{\pi} \frac{d^2N}{dp_T^2dy} 
\left[
1 + 2 \sum_{n=1}^{\infty} v_n(p_T,y) \cos(n\phi)
\right]~,
\end{equation}
are called {\it directed\/} and {\it elliptic\/} flow. Here, 
$p_T=(p_x^2+p_y^2)^{1/2}$ is the transverse momentum, $y$ the 
rapidity and $\phi$ the azimuthal angle between the particle's 
momentum and the reaction plane. While the elliptic flow 
$v_2 = \langle \cos(2\phi) \rangle = 
\langle (p_x/p_T)^2 - (p_y/p_T)^2 \rangle$ 
is assumed to be particularly sensitive to the pressure at the 
early stages of the collisions 
\cite{Kolb:2000fh,ell_flow,Sorprl97,Olli92+98}, 
the directed flow appears to develop until the late stage of the 
heavy-ion reaction \cite{DF_prc00,DF_prc01,LPX99}. On the other hand, 
the directed flow of particles with high transverse momentum, which 
are emitted at the onset of the collective expansion, can carry 
information about the EOS of the dense matter phase from the initial 
conditions. 
A lot of measurements of the $v_2$ parameter in Au+Au collisions 
at RHIC have been performed for charged and identified hadrons, 
see \cite{ellST01,PHOB_fl,BackPHOBOS:2004,Adler2003} and references 
therein, whereas experimental results for the directed flow $v_1$ 
at RHIC has been reported only quite recently 
\cite{Adams:2003zg,BeltTonjes:2004jw}.


\subsection{Directed flow of inclusive charged hadrons}
\label{dirflow}

\begin{figure}[htb]
\includegraphics[scale=0.5]{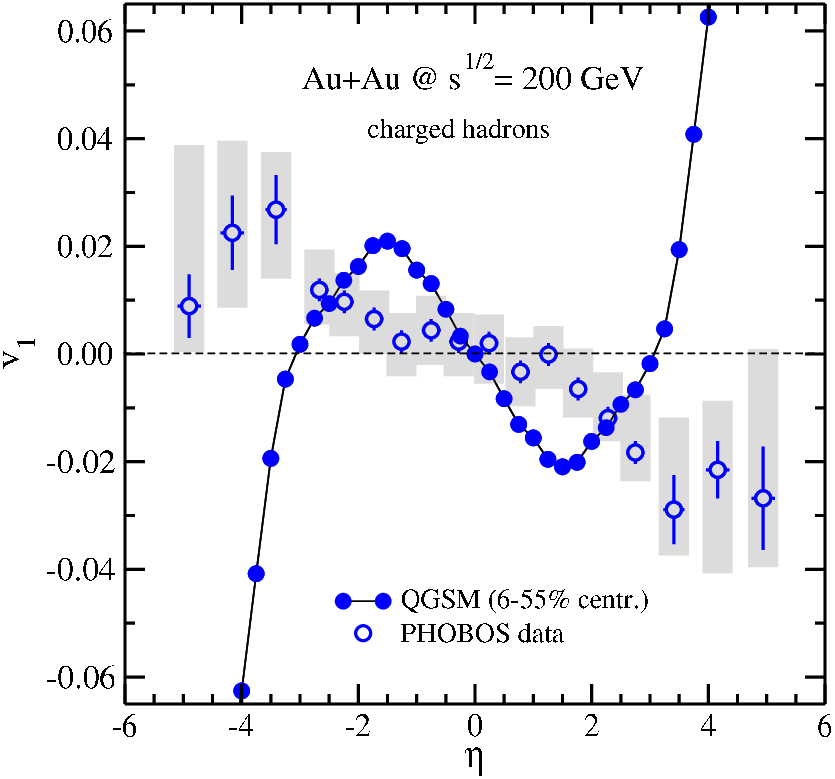}
\caption{Directed flow $v_1$ for charged hadrons as a function of 
pseudorapidity $\eta$ in comparison to the result from the 
PHOBOS Collaboration (centrality 6\% to 55\%) \cite{BeltTonjes:2004jw} 
for Au+Au collisions at $\sqrt{s_{NN}} = 200~{\rm GeV}$. 
Statistical (bars) and systematic (boxes) experimental errors are shown.
\label{fig:v1etahchar1}}
\end{figure}

Figure~\ref{fig:v1etahchar1} depicts the QGSM simulation result for 
the pseudorapidity dependence of the directed flow $v_1(\eta)$ of 
charged hadrons compared to the experimental data from the 
PHOBOS Collaboration \cite{BeltTonjes:2004jw} for 6\% to 55\% central 
Au+Au collisions at $\sqrt{s_{NN}} = 200~{\rm GeV}$ (QGSM predictions 
for the directed flow of identified particles at lower RHIC energy, 
$\sqrt{s_{NN}} = 130~{\rm GeV}$, can be found in 
\cite{qgsm_dir_fl_130}).
\begin{figure*}[htb]
\includegraphics[scale=0.525]{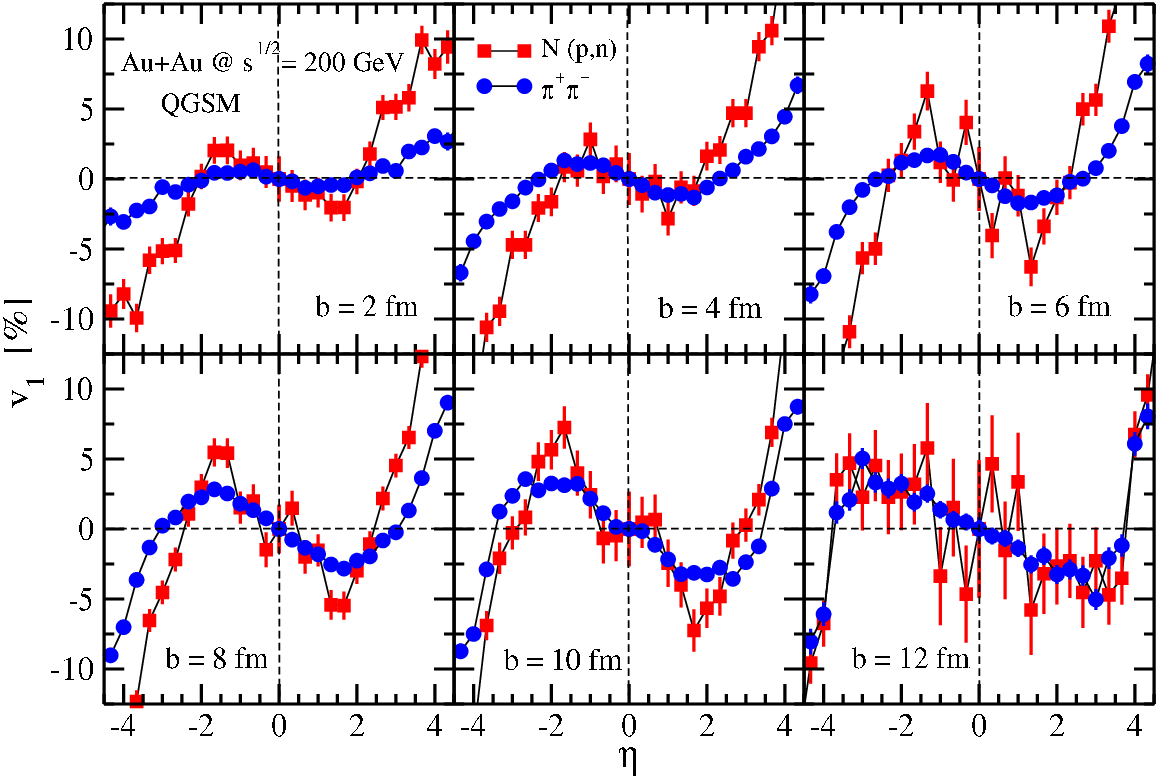}
\caption{Directed flow $v_1$ for nucleons and charged pions from 
$\sqrt{s_{NN}} = 200~{\rm GeV}$ Au+Au reactions as a function of 
pseudorapidity $\eta$ for six different impact parameters $b$ 
going from central ($b=2~{\rm fm}$) to peripheral ($b=12~{\rm fm}$) 
collisions. Only statistical errors are shown.
\label{fig:v1etahchar2}}
\end{figure*}
One can see that the model reproduces the $v_1$ data quite well both
qualitatively and quantitatively, although the maxima of the directed 
flow around $|\eta| \approx 2$ are shifted to lower 
pseudorapidities compared to the experimental data. 
The $v_1(\eta)$ result show a characteristic {\it wiggle\/} 
structure with a clear antiflow\footnote{Conventionally, the type of 
flow with positive slope $d v_1/d \eta$ is called {\it normal flow\/}, 
in contrast to {\it antiflow\/} for which $d v_1/d \eta$ is negative.} 
component in the middle $|\eta|$ region. 
It was pointed out in \cite{Snellings:1999bt} (see also
\cite{DF_prc01,Zabrodin:2004gi}) that the phenomenon leading to the 
formation of a {\it wiggle\/} structure for the directed flow of
nucleons is caused by dense baryon-rich matter shadowing, 
which plays a decisive role in the competition between normal flow 
and antiflow in noncentral nuclear collisions at ultra-relativistic 
energies. Within microscopic string model calculations such deviations 
from the straight line behaviour of the nucleonic flow were first 
observed in very peripheral Au+Au collisions at AGS energy \cite{Br95}. 
Experimentally the {\it wiggle\/} structure of $v_1$ for protons in 
peripheral Pb+Pb collisions at SPS has been observed by the NA49 
Collaboration \cite{Wetzler:2002fi,Alt:2003ab}. 
However, the QGSM distributions of $v_1(\eta)$ at these lower energies 
also have peaks which are shifted by approximately one unit of rapidity 
toward $\eta = 0$ \cite{Am91,DF_prc00,DF_prc01} compared to the 
experimental data. So, this shift, which also leads to a steeper slope 
in the mid-rapidity region, seems to be a sort of model `artefact' 
occurring at higher collision energies as well. 
The question about the origin of the directed flow's rapidity dependence 
obtained within the QGSM is not so easily to answer. One reason might be 
the lack of heavy resonances in the model compared, e.g., to the 
Ultra-Relativistic Quantum Molecular Dynamics (UrQMD) model \cite{BlSt02} 
or the multi-phase transport (AMPT) model \cite{Chen:2004vh}. 
These resonances are mostly concentrated in the rapidity regions of 
flying away spectators which exhibit normal flow. Lighter particles 
coming from the decays of the heavy resonances, each moving in the 
normal direction and having nearly the same rapidity, may significantly 
enhance the normal component of the directed flow in a pseudorapidity range 
closer to the fragmentation region. The total multiplicity of particles 
with pseudorapidity $|\eta| > 3.5$ is quite low in the present version of 
the QGSM. 
The complex connection between the model dynamics of the QGSM and the 
characteristics of the resulting directed flow requires further lucid 
investigations to understand in detail the origin of the change in sign, 
and in particular the strong antiflow behaviour of $v_1(\eta)$ in the 
mid-rapidity region.

Nevertheless, our picture is a bit different compared to that provided 
by the microscopic models based on the FRITIOF routine. For instance, 
the relativistic quantum molecular dynamics (RQMD) model favours weak
but still normal flow for pions \cite{Snellings:1999bt} even for
more peripheral topologies with $b = 5-10$ fm, corresponding to 
a centrality range $\sigma/\sigma_{geo} = 15-60\%$. 
Both the UrQMD model and the AMPT model show a very flat and essentially 
zero directed flow \cite{BlSt02,Chen:2004vh} in a broad range 
$|\eta| \leq 2.5$. -- It is worth mentioning here that the results
of both models have been obtained for minimum bias, not 
semi-peripheral, events. -- Although the data seem to indicate 
antiflow behaviour for the directed flow of charged particles with 
the possible flatness at $|\eta| \leq 1.5$, the measured signal is 
quite weak, -- the magnitude of the flow is less than 1\% at $|\eta| 
\leq 2$. Therefore, relatively large systematic error bars do not
permit to disentangle between the different models.

Similar antiflow alignment can be obtained also within the multi
module model (MMM) \cite{MCS02}, which is based on fluid dynamics 
coupled to the formation of colour ropes. In this model the effect 
of the tilted initial state, responsible for the antiflow formation,
reaches its maximum for the impact parameter $b \approx 0.5 (R_A +
R_B)$, i.e. for the centrality $\sigma/\sigma_{geo} \approx 25\%$
in case of a symmetric system of colliding nuclei. To check the
centrality dependence of the directed flow at full RHIC energy,
the pseudorapidity distributions $v_1(\eta)$ of nucleons and charged 
pions in Au+Au collisions at $\sqrt{s_{NN}} = 200~{\rm GeV}$ are 
displayed in Fig.~\ref{fig:v1etahchar2} for six different impact 
parameters $b$ going from central (upper left panel) to very 
peripheral (lower right panel) configurations. 
Regardless of the centrality range, directed flow of pions has 
negative slope, i.e. antiflow, in the mid-rapidity range. For nucleons
the azimuthal anisotropy parameter $v_1$ is small for central 
collisions ($b=2~{\rm fm}$). In accord with our previous studies 
and conclusions \cite{DF_prc00,DF_prc01}, for Au+Au collisions at full 
RHIC energy deviations of nucleonic flow from a normal flow behaviour 
occur already at quite small impact parameters. This means that the
effect is indeed shifted to more central configurations. With the
increase of $b$ these deviations, representing the {\it wiggle\/} 
structure of the flow, appear more distinctly. 
However, the QGSM simulations for both nucleonic and pionic flow at 
mid-rapidity $|\eta| \leq 1$ are consistent with the zero flow signal 
$v_1 = 0$. The other features which should be mentioned here are 
the broadening of the antiflow region and the increase of its strength 
as the reaction becomes more peripheral. Therefore, one can disentangle 
between two processes of different origin employed for the description 
of the {\it third-flow component\/}: If the formation of nucleonic 
antiflow is dominated by the creation of QGP, the flow maximum is 
reached at $b \leq 6$~fm \cite{MCS02}, whereas for the shadowing scenario 
the strong antiflow should be observed also in very peripheral events 
with $b \approx 10$~fm.


\subsection{Elliptic flow of charged and identified hadrons}
\label{ellflow}

\begin{figure}[htb]
\includegraphics[scale=0.5]{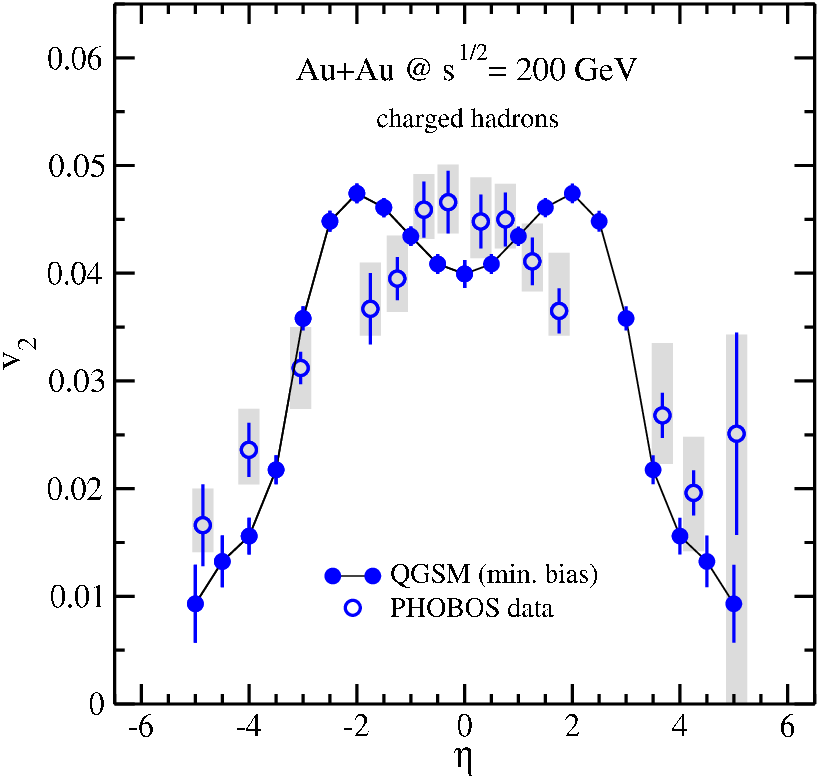}
\caption{Elliptic flow $v_2$ for inclusive charged hadrons as a function 
of pseudorapidity $\eta$ in comparison to the PHOBOS data of minimum 
bias Au+Au collisions at $\sqrt{s_{NN}} = 200~{\rm GeV}$ 
\cite{BackPHOBOS:2004}. The systematic errors of the experimental data
are shown as gray boxes together with the statistical errors (bars), 
respectively.
\label{fig:v2etahchar1}}
\end{figure}

Here we investigate the pseudorapidity dependence 
of the elliptic flow of charged hadrons. The QGSM simulation result 
for minimum bias Au+Au collisions at $\sqrt{s_{NN}} = 200~{\rm GeV}$ 
are compared in Fig.~\ref{fig:v2etahchar1} with the experimental data 
of the PHOBOS Collaboration \cite{BackPHOBOS:2004}. 
The elliptic flow displays a strong in-plane alignment in accordance
with the predictions of Ref.~\cite{Olli92+98}. 
At mid-rapidity $|\eta| < 1.0$ the elliptic flow is almost constant. 
Then it rises up slightly and drops rapidly at $|\eta| > 2.0$ with 
increasing pseudorapidity. The mean value of 
$v_2^{\rm ch}(|\eta| \le 2.5)$ is practically as large as the value
measured by the PHOBOS Collaboration at mid-rapidity. 
The experimental data, which indicate a steady decrease in $v_2$
with increasing $|\eta|$, are slightly overestimated by the model 
only at $|\eta| \approx 2.0$ as a consequence of a double hump 
structure in the theoretical result. This difference in shape 
close to $\eta = 0$ still rankles somewhat, although within the error 
bars the pseudorapidity dependence of the elliptic flow of charged 
hadrons obtained within the QGSM shows a really fair agreement with 
the PHOBOS data \cite{BackPHOBOS:2004} in the whole $\eta$ range.
\begin{figure*}[htb]
\includegraphics[scale=0.525]{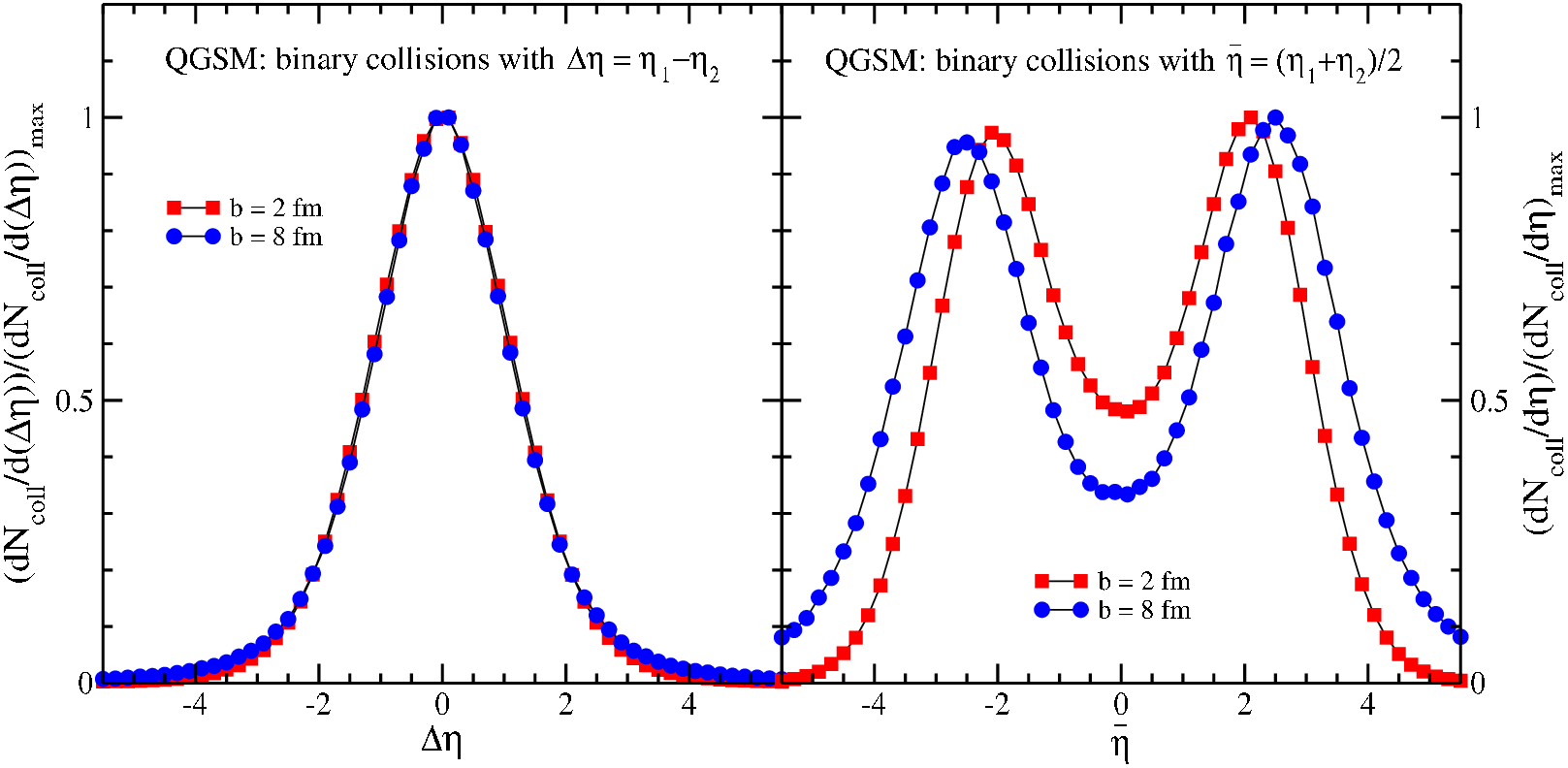}
\caption{The number of binary particle collisions per pseudorapidity 
interval normalized to its maximum as function of the pseudorapidity 
difference $\Delta \eta = \eta_1 - \eta_2$ (left) and the mean 
pseudorapidity $\bar{\eta} = (\eta_1 + \eta_2)/2$ (right) of the 
two colliding particles. The QGSM results are obtained for simulated 
Au+Au reactions at $\sqrt{s_{NN}} = 200~{\rm GeV}$ with two different 
impact parameters $b$.
\label{fig:etaNcoll}}
\end{figure*}
%

However, following the idea of longitudinal boost invariance\footnote{The 
assumption of a longitudinal boost invariant system means that its 
energy density and pressure do not depend on the longitudinal 
coordinate $z$ compared at the same proper time $\tau = \sqrt{t^2 - z^2}$. 
In other words, the evolution of the pressure and energy density 
depends only on $\tau$ but not on $\eta$.} of the expanding hot and 
dense matter and the common interpretation of elliptic flow as a 
consequence of secondary particle collisions, one would expect no 
or at least a weak pseudorapidity dependence of $v_2$ similar to that 
of the multiplicity density $dN/d\eta$ over a large rapidity range. 
The experimentally observed multiplicity stays approximately constant 
within three units of pseudorapidity \cite{Back:2002wb}, while the 
elliptic flow data show a pronounced peak at mid-rapidity 
\cite{BackPHOBOS:2004,Back:2004je}. This is somehow in contradiction 
with the assumption of longitudinal boost invariance over a broad 
region of rapidity in RHIC collisions. 
Most of the hydrodynamics calculations reported in the literature 
are based on boost invariant models. 
Therefore the results obtained within such approaches are 
independent of rapidity and one is limited to discuss only the 
transverse behaviour \cite{Huovinen:2003fa}. 
Hydrodynamics results for the pseudorapidity dependence of $v_2$ 
are scarce. The shape of $v_2(\eta)$ in Au+Au collisions at 
$\sqrt{s_{NN}} = 130~{\rm GeV}$ studied within a full 
three-dimensional hydrodynamic model by Hirano {\it et al.} 
\cite{Hirano:2002hv,Hirano:2002ds,Hirano:2001eu} shows a bump 
structure in forward and backward rapidity regions, when the 
initial energy density profile depends strongly on the space-time 
rapidity $\eta_s$. The bumps appear at the same rapidity regions 
where the initial configuration is not boost invariant anymore. 
In the case of an almost $\eta_s$ independent deformation of the 
initial energy density, the resultant $v_2(\eta)$ has no bumps. 
Thus it was shown in \cite{Hirano:2001eu} that the pseudorapidity 
dependence of the elliptic flow is highly sensitive to the 
parametrization of the initial energy density profile in the 
longitudinal direction. 
The initial energy density profile in microscopic transport models 
is not explicitly parametrized, but it is implicitly fixed by the 
initial conditions of the projectile and target nucleus. 
The elliptic flow studied within the UrQMD model in the cascade mode 
shows also a prominent dip at central rapidities for all inspected 
hadrons \cite{BlSt02}. There it is argued that this rapidity behaviour 
of $v_2$ indicates a region of small interaction strength (or low 
`pressure') because of the direct connection between the strength 
of the anisotropic flow and the mean free path of the particles 
forming the hot mid-rapidity region. Therefore, the appearance of 
the dip in $v_2$ is linked to a feature of the model dynamics in 
the early stage, namely the pre-equilibrium string dynamics and 
interactions on the parton level \cite{BlSt02}. 
Also in the QGSM, the emergence of the double bump structure in 
$v_2(\eta)$ is strongly connected with the model dynamics, as one can 
clearly see in the time evolution of the elliptic flow and its 
rapidity dependence \cite{Bravina:2004td}. The normalized number 
of binary particle collisions per pseudorapidity interval throughout 
the evolution of the system as a function of the pseudorapidity is 
depicted in Fig.~\ref{fig:etaNcoll} for two impact parameters 
$b = 2~{\rm fm}$ and $b = 8~{\rm fm}$ to show the strong correlation 
between the particle interactions and the flow strength mainly 
produced by secondary collisions. 
First of all, this number versus the pseudorapidity difference 
$\Delta \eta = \eta_1 - \eta_2$ of the two colliding particles 
is plotted in the left panel of Fig.~\ref{fig:etaNcoll}. 
The peak around $\Delta \eta = 0$ indicates unambiguously that 
independently of the impact parameter the particles with equal or 
at least very similar rapidities interact most likely. 
The plot on the right hand side of Fig.~\ref{fig:etaNcoll}, where 
the number of binary collisions normalized to its maximum is shown 
as a function of the mean pseudorapidity $\bar{\eta} = (\eta_1 + \eta_2)/2$ 
of the two colliding particles, is even more instructive. 
Here, a clear double peak structure appears in the same rapidity 
region as seen in Fig.~\ref{fig:v2etahchar1} for the $\eta$ dependence 
of the elliptic flow. The dip in the number of collisions around 
$\bar{\eta} = 0$ becomes more pronounced with decreasing centrality. 
In addition, the two peaks for $b = 8~{\rm fm}$ are slightly shifted 
to higher pseudorapidities compared with the result for $b = 2~{\rm fm}$. 
This is nicely consistent with the centrality dependence of $v_2(\eta)$ 
obtained within the QGSM as it will be discussed in the following.

%
\begin{figure}[htb]
\includegraphics[scale=0.5]{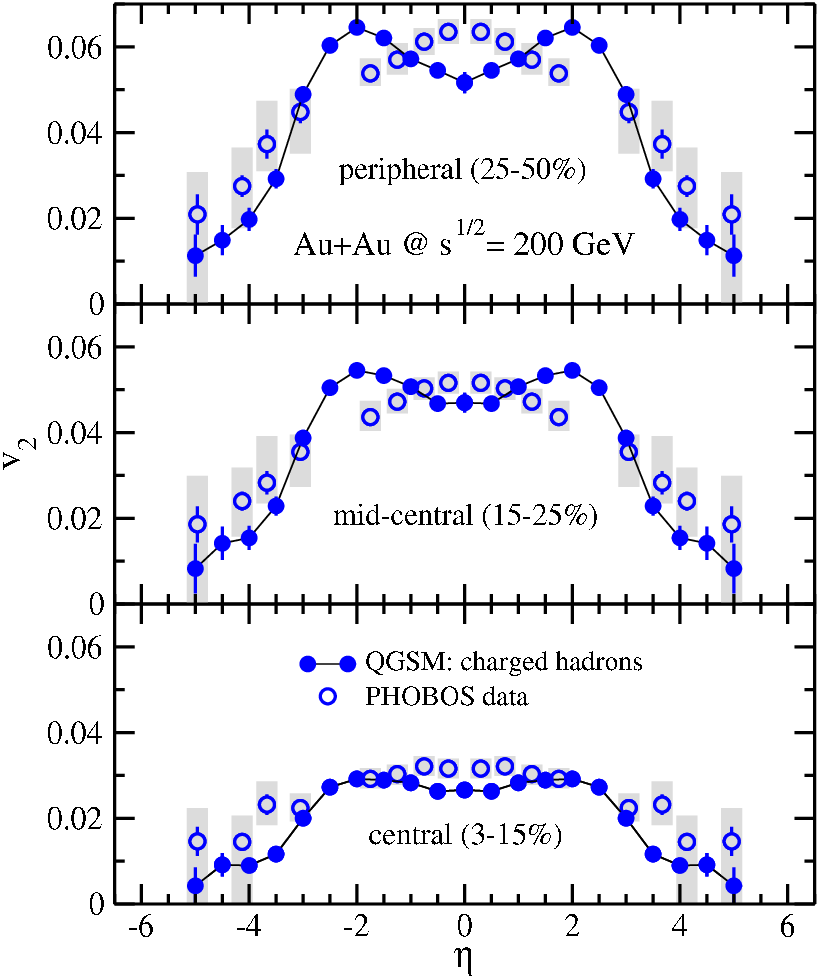}
\caption{Elliptic flow $v_2$ for inclusive charged hadrons from 
$\sqrt{s_{NN}} = 200~{\rm GeV}$ Au+Au collisions as a function of 
pseudorapidity $\eta$ for the three centrality classes according 
to the PHOBOS analysis (combined hit- and track-based results) 
\cite{BackPHOBOS:2004}. 
Again, the systematic errors of the experimental data are shown as 
gray boxes together with the statistical error bars, respectively.
\label{fig:v2etahchar2}}
\end{figure}
Figure~\ref{fig:v2etahchar2} presents the pseudorapidity distribution 
of the elliptic flow for charged hadrons in gold-gold collisions 
at full RHIC energy for three different centrality classes, ranging 
from central (bottom panel) via mid-central (middle panel) to 
peripheral (top panel) in accordance with the definitions in 
\cite{BackPHOBOS:2004}. 
The results from the QGSM simulation and the PHOBOS analysis 
(combined data from the hit- and track-based methods) are overlaid. 
The model is able to describe the magnitude and shape of 
$v_2^{\rm ch}(\eta)$ quite well across all of the three centrality 
bins within the given systematic and statistical uncertainties. 
The overall shape of the $\eta$ distribution changes only very little 
with centrality and shows a behaviour very similar to that depicted 
in Fig.~\ref{fig:v2etahchar1} for the minimum bias events. 
The two-peak structure of $v_2^{\rm ch}(\eta)$ in the interval 
$|\eta| \le 2.5$, which is clearly seen for simulated peripheral 
collisions but `washed out' in the central bin, is not attributed 
solely to QGSM, but arises also in e.g. UrQMD calculations 
\cite{BlSt02} as discussed in detail above. However, the effect is 
small, and the mean value of the $v_2^{\rm ch}$ parameter over the 
aforementioned range increases from central to peripheral collisions 
in good quantitative agreement with the experimental data. This is a 
not so trivial result, because neither the pseudorapidity nor the 
centrality dependence of the elliptic flow, which is discussed below, 
is reproduced correctly at RHIC so far by the UrQMD calculations 
and hydrodynamic models.

\begin{figure}[htb]
\includegraphics[scale=0.5]{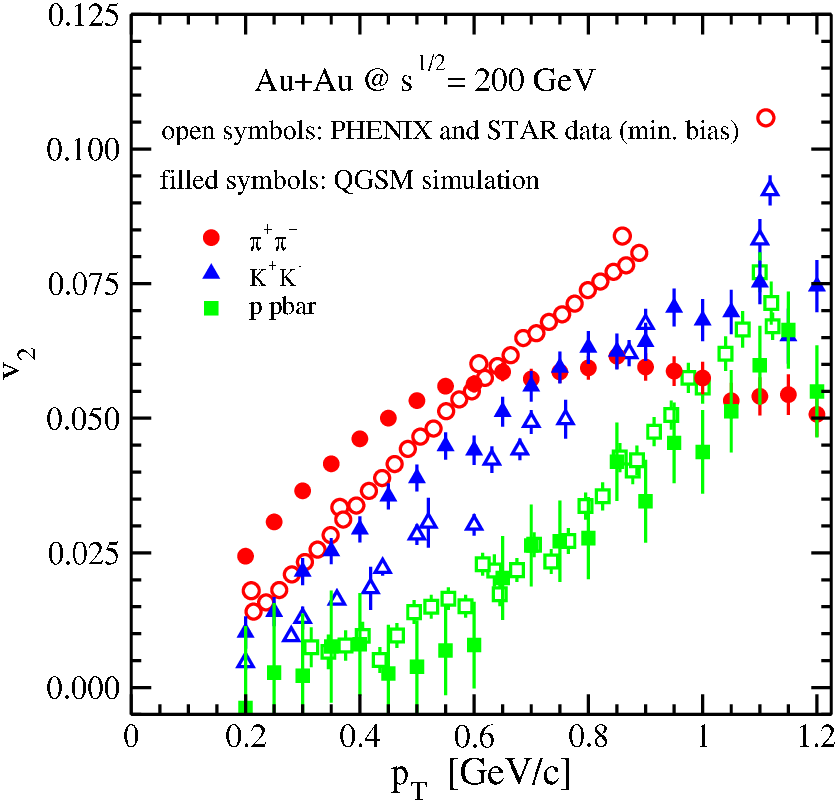}
\caption{Transverse momentum dependence of $v_2$ for identified 
hadrons $\pi^+ + \pi^-$, $K^+ + K^-$, $p + \bar{p}$ in the low-$p_T$ 
range $0.0 \le p_T \le 1.0~{\rm GeV/}c$ for minimum bias Au+Au 
collisions at $\sqrt{s_{NN}} = 200$ GeV.
The experimental data are PHENIX \cite{Adler2003} and 
STAR \cite{Adams:2004bi} results. Only statistical errors are shown.
\label{fig:v2pThcode1}}
\end{figure}

\begin{figure}[htb]
\includegraphics[scale=0.5]{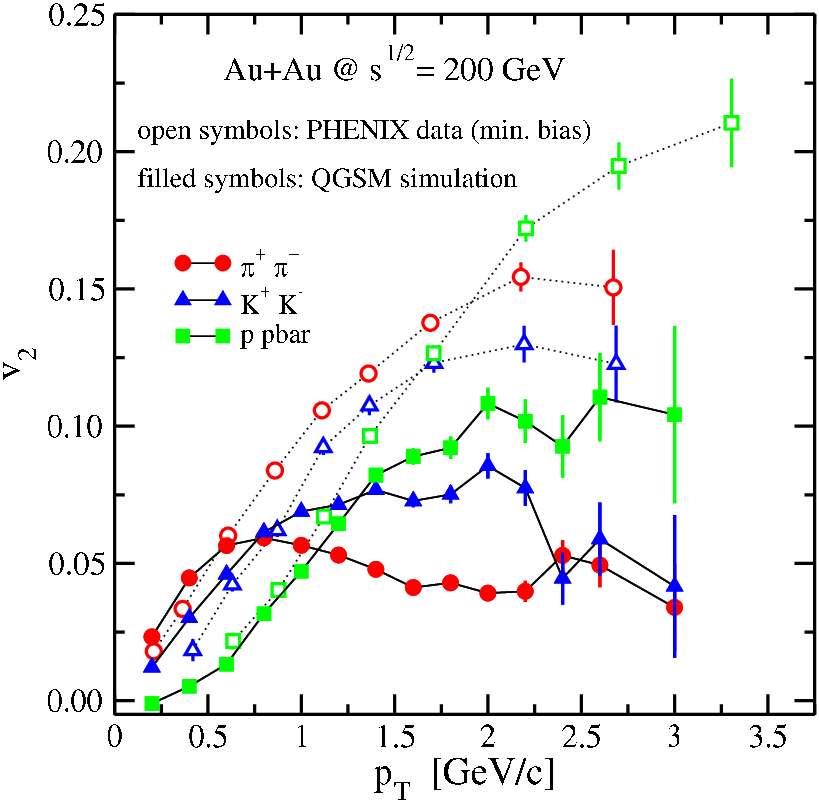}
\caption{The same like in Figure~\ref{fig:v2pThcode1} but in the 
overall $p_T$ range $0.0 \le p_T \le 3.5~{\rm GeV/}c$. The PHENIX 
data are taken from \cite{Adler2003}. Again, only statistical errors 
are shown.
\label{fig:v2pThcode2}}
\end{figure}

The next observable is the transverse momentum dependence of $v_2$ for 
identified hadrons, namely, combined $\pi^+ + \pi^-$, $K^+ + K^-$,
as well as $p + \bar{p}$ spectra in minimum bias Au+Au collisions at 
$\sqrt{s_{NN}} = 200~{\rm GeV}$. The QGSM simulation results are 
depicted in Figs.~\ref{fig:v2pThcode1} and \ref{fig:v2pThcode2} 
in comparison to the PHENIX \cite{Adler2003} and STAR data 
\cite{Adams:2004bi}, respectively.
The agreement with the experimental data at least for 
protons/antiprotons and kaons is rather good in the range of low 
transverse momenta $0.0 \le p_T \le 1.0~{\rm GeV/}c$, as illustrated in 
Fig.~\ref{fig:v2pThcode1}. The steady increase of the elliptic flow 
with rising $p_T$ and the larger $v_2$ parameter at a given $p_T$ 
for the lighter mass hadrons compared to the heavier ones is well 
reproduced by the model calculation. For pions the latter statement 
is only valid for the transverse momenta below $0.75~{\rm GeV/}c$. 
The elliptic flow of charged pions starts already to saturate at 
$p_T > 0.5~{\rm GeV/}c$. 
This early saturation of the pionic $v_2$ continues with a slight 
decreasing at increasing $p_T$ and entails the peculiar result 
to be smaller than the kaon flow for transverse momenta between 
$1.0~{\rm GeV/}c$ and $2.5~{\rm GeV/}c$, as shown for the 
overall $p_T$ range in Fig.~\ref{fig:v2pThcode2}. 

\begin{figure}[htb]
\includegraphics[scale=0.5]{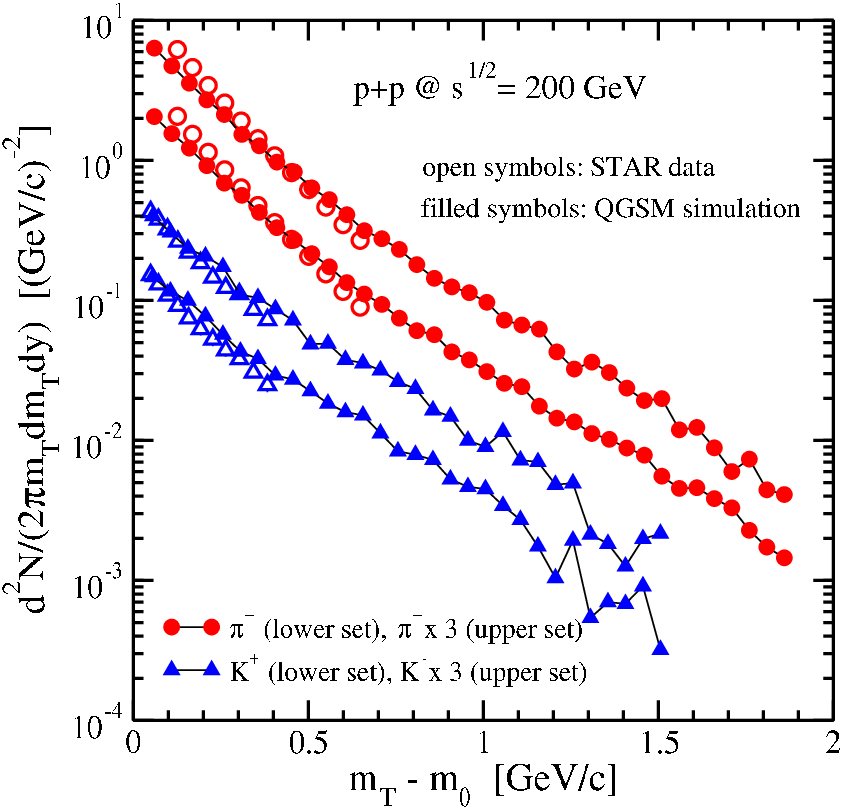}
\caption{Transverse mass spectra of $\pi^{\pm}$ and $K^{\pm}$ 
for simulated $pp$ collisions at $\sqrt{s_{NN}} = 200~{\rm GeV}$ 
compared to the STAR data \cite{AdamsSTAR:2004}.
\label{fig:mTspectra}}
\end{figure}

This deviation compared to the experimental findings is remarkable 
because it was shown in \cite{ell_flow} that the magnitude of the 
pionic flow in the QGSM calculations is already twice as large as 
obtained, e.g. in the RQMD ones. It has been elaborated in 
\cite{ell_flow} that the contributions of hard processes and 
multi-Pomeron exchanges are very important to reach the reported 
magnitude of $v_2$ and to reproduce the particle multiplicities 
correctly. To confirm the latter statement, the transverse mass 
spectra of positively and negatively charged pions and kaons for 
$pp$ collisions at $\sqrt{s_{NN}} = 200~{\rm GeV}$ are presented in 
Fig.~\ref{fig:mTspectra}. 
The $m_T$ spectra obtained within the full version of the QGSM, which 
incorporates hard and multichain contributions, are in good agreement 
with the experimental spectra reported by the STAR Collaboration 
\cite{AdamsSTAR:2004}.

Coming back to the $p_T$ behaviour of the elliptic flow depicted in 
Fig.~\ref{fig:v2pThcode2} it is important to stress at this point 
another striking feature of the QGSM. Namely, this model is able to 
reproduce at least qualitatively the experimental evidence of the 
crossing of the elliptic flow for mesons and baryons at 
$p_T \approx 1.7~{\rm GeV/}c$ \cite{Adler2003}. 
The simulation data show that at $p_T < 1.4~{\rm GeV/}c$ the $v_2$ 
parameter for kaons is larger than for (anti)protons. At higher 
transverse momenta $p_T > 1.4~{\rm GeV/}c$ the situation is completely 
changed. Here, the elliptic flow of protons and antiprotons becomes 
larger than $v_2$ for charged kaons and pions.

\begin{figure}[htb]
\includegraphics[scale=0.5]{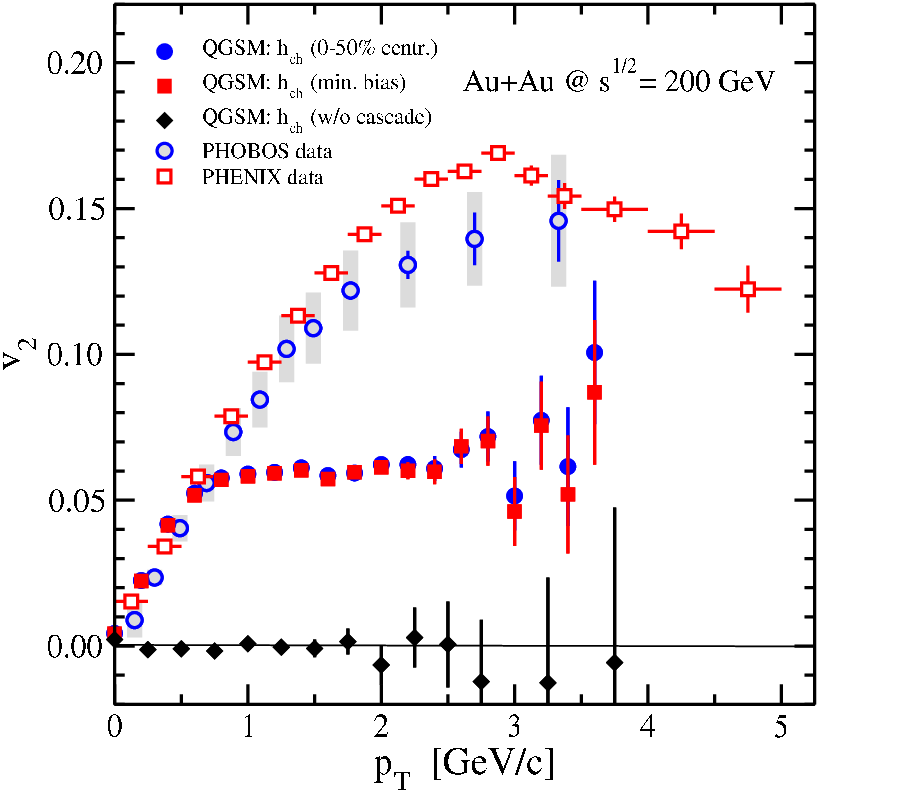}
\caption{Elliptic flow $v_2$ for inclusive charged hadrons as a 
function of the transverse momentum $p_T$ in comparison to PHOBOS 
\cite{BackPHOBOS:2004} and PHENIX \cite{Adler2003} data. 
Note the different centrality ranges of the data sets. 
All shown errors are statistical.
\label{fig:v2pThchar}}
\end{figure}

This behaviour is also described within the quark recombination 
models \cite{MV03,GKL2003,FMNB2003,HY2003}, which assume the 
statistical coalescence of two or three quarks into a hadron. In 
contrast, the hydrodynamic picture shows the same mass-ordering for 
the elliptic flow of different particles at all transverse momenta 
\cite{HKHRV2001}. We will come to this point after the study of 
$v_2(p_T)$ for inclusive charged hadrons. The QGSM distributions 
are shown in Fig.~\ref{fig:v2pThchar} in comparison to the 
experimental data of the PHOBOS and PHENIX Collaborations 
\cite{BackPHOBOS:2004,Adler2003}. 
To demonstrate the importance of rescattering processes for 
the elliptic flow formation the $v_2^{\rm ch}$ extracted from a 
QGSM simulation run for minimum bias Au+Au collisions without the 
hadronic cascade is additionally plotted in this figure.
It is quite obvious that the model without subsequent secondary 
interactions of produced hadrons creates zero elliptic flow, 
because the azimuthal distributions of secondaries in elementary
hadron-hadron collisions are isotropic. The anisotropic flow, i.e. 
the azimuthal anisotropies of the number of produced hadrons, 
results from the spatial asymmetry of the collision zone and 
subsequent rescattering processes, which are crucial to convert 
the initial spatial anisotropy into the final momentum anisotropy. 
Hadronic rescattering including hard and multichain contributions 
creates an elliptic flow, which rises almost linearly according to 
the experimental data within the interval 
$0.0 \le p_T \le 1.0~{\rm GeV/}c$, but saturates already for 
transverse momenta above $1.0~{\rm GeV/}c$ at a level of 
$v_2^{\rm ch} \approx 6\%$ whereas the experimentally measured flow 
increases further up to $v_2^{\rm ch} \approx 14-16\%$ ere it 
declines for $p_T \ge 3.0~{\rm GeV/}c$. 
Although the model seems to describe the $p_T$ dependence 
qualitatively well, it underestimates the experimental elliptic flow 
of charged particles with transverse momenta above $1.0~{\rm GeV/}c$ 
roughly by up to $50\%$. 
What is the reason of these deviations? Recall that hadrons in the 
QGSM gain the transverse momentum due to (i) transverse motion of 
the constituent quarks and (ii) transverse momentum of constituents
acquired in the course of string fragmentation. The parameters of 
these two processes are fixed by comparison with the available 
hadronic data. Other sources of the transverse motion are (iii) the
transverse Fermi motion of nucleons in the colliding nuclei and (iv)
rescattering of produced particles in the hot and dense nuclear medium.
It was already mentioned that the latter process is the most crucial
for the development of the elliptic flow in nuclear reactions.  
However, secondary hadrons with high transverse momenta experience
in average less collisions than their low-momentum counterparts 
because of the large formation time (which originates from the 
uncertainty principle). 
As was shown in \cite{BlSt02}, in the limit of vanishing formation time 
the elliptic flow increases drastically. But all parameters linked to 
the formation time of produced particles in the QGSM are also fixed by
comparison with experimental data on hadronic interactions. Therefore,
the presented elliptic flow can be considered as an upper limit 
obtained within the hadronic cascade scenario. It clearly indicates on 
new physical effects not taken into account by the microscopic model. 
Jets are among the most likely candidates for these processes. Indeed, 
as was discussed in \cite{LoSn01}, the non-uniform dependence of the 
energy loss on the azimuthal angle results in azimuthal anisotropy of 
jet spectra in non-central nuclear collisions. This leads to a 
significant increase of the elliptic flow of high-$p_T$ particles.
Another possible explanation is, e.g. a dramatic increase of all 
$s$-channel transition rates in the vicinity of the chiral phase 
transition \cite{GBA04}, which causes a critical opacity and fast 
thermalization in the system.

\begin{figure*}[tb]
\includegraphics[scale=0.525]{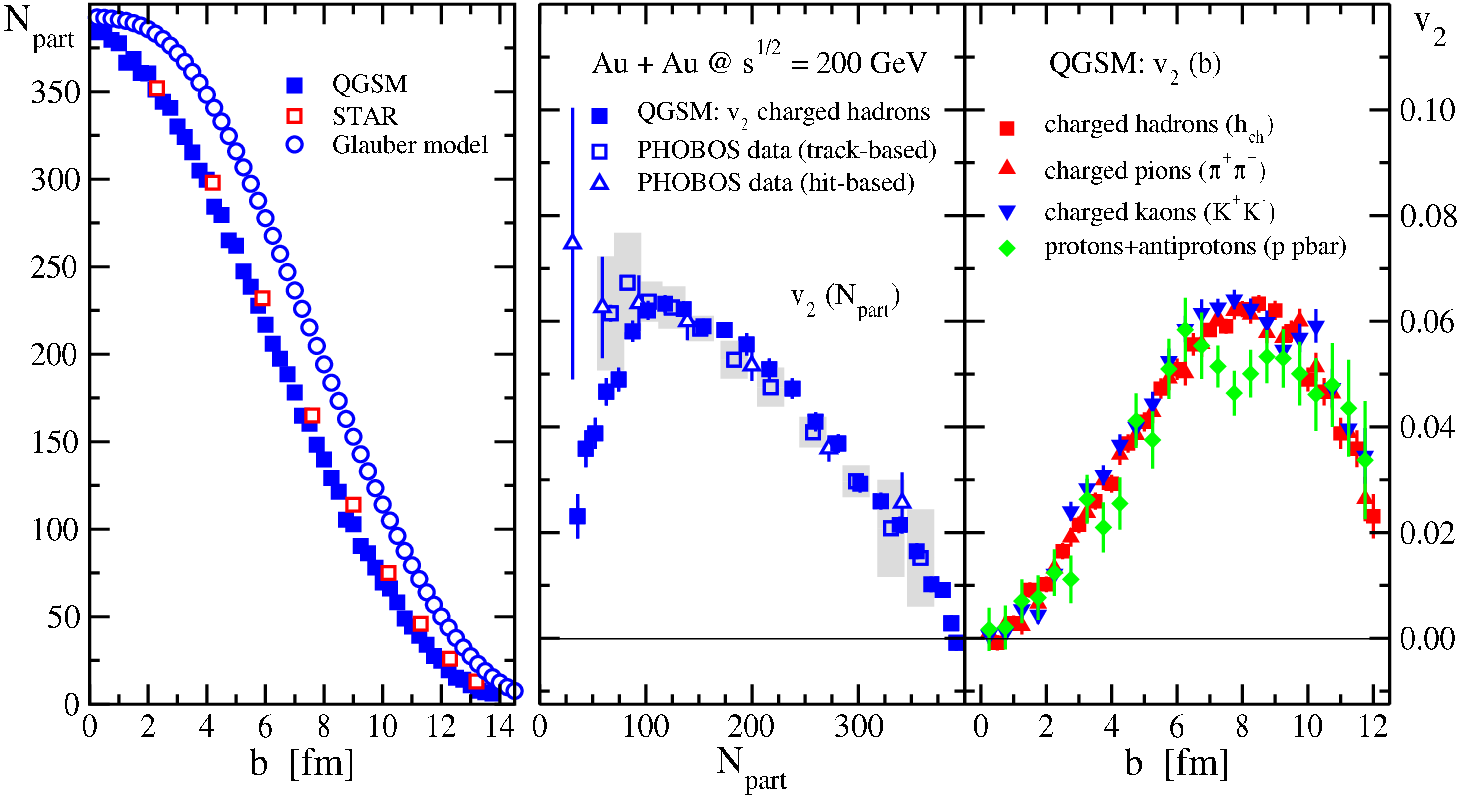}
\caption{Centrality dependence of the elliptic flow $v_2$ for charged 
hadrons compared with the PHOBOS data \cite{BackPHOBOS:2004} (middle 
panel), and $v_2$ of inclusive as well as identified charged hadrons 
as a function of the impact parameter $b$ (right panel). 
The $b$ dependence of the number of participants $N_{\rm part}$ 
is shown in the left panel (the STAR data are taken from \cite{Adams:2004bi}). 
The errors are statistical.
\label{fig:v2centrhchar}}
\end{figure*}
Finally, the centrality dependence of the total, i.e. $\eta$ and 
$p_T$ integrated $v_2$ for charged hadrons is depicted in the middle 
and right panel of Figure~\ref{fig:v2centrhchar}. Since the centrality 
of collision events in the experimental measurements \cite{Adler2003} 
is characterised by the mean number of participants $N_{\rm part}$ 
(seen in the middle panel), 
we show in addition to this signal the original impact parameter 
dependence of the elliptic flow (right panel). 
The relation between the number of participants $N_{\rm part}$
in the Au+Au reactions and the impact parameter $b$ of the simulated 
collisions is shown in the left panel of Figure~\ref{fig:v2centrhchar}. 
The number of nucleons participating in inelastic scatterings, i.e. 
$N_{\rm part}$, is directly available in the QGSM. This number depends 
mainly on the nucleon-nucleon cross section, which for the given 
initial energy of $\sqrt{s_{NN}} = 200~{\rm GeV}$ is calculated 
in the framework of GRT. Apart from the cross section and its energy 
dependence, the number of participants varies with the impact parameter 
like $\propto \exp(-b^2)$. 
Thus, the $b$-dependence of $N_{\rm part}$ within the QGSM is rather 
different compared to the Glauber model result, but in nice agreement 
with the experimentally estimated  $N_{\rm part}(b)$ \cite{Adams:2004bi}. 
Amazingly, the elliptic flow of inclusive charged hadrons extracted 
from our simulation almost coincides with the PHOBOS experimental data. 
This remarkable result in conjunction with the transverse momentum 
dependence of $v_2^{\rm ch}$ shown in Fig.~\ref{fig:v2pThchar} reflects 
the dominance of hadrons with low $p_T$. 
Furthermore one can see that the elliptic flow as a function of the 
impact parameter $b$ reaches a maximal value of around 6\% at 
$b \approx 8$~fm, corresponding to $N_{\rm part} \approx 100$, and 
decreases with further increasing $b$. As expected, the flow in the 
mid-rapidity region is caused mainly by pions, but the azimuthal 
anisotropy parameters $v_2(b)$ for charged kaons and combined 
protons/antiprotons are also very similar within the statistical 
uncertainties. Note that the magnitude of the total pionic flow 
obtained within the QGSM simulations is more than twice as large as 
created by the RQMD ones \cite{Snellings:1999gq}.


\section{Summary and conclusions}
\label{sum}

In summary, the pseudorapidity distributions of the azimuthal 
anisotropy parameters $v_1(\eta)$ and $v_2(\eta)$ of inclusive 
charged hadrons and their centrality dependence has been studied in 
Au+Au collisions at full RHIC energy of $\sqrt{s_{NN}}=200~{\rm GeV}$ 
within the microscopic quark-gluon string model. The QGSM simulation 
results for the directed flow show antiflow alignment within the 
pseudorapidity range $|\eta| \leq 2$ in a fair agreement with the 
experimental $v_1(\eta)$ data, but cannot reproduce the further 
development of the antiflow up to $|\eta| \approx 3.5$.
In a broad pseudorapidity region the model generates a {\it wiggle\/} 
structure for the directed flow of nucleons $v_1^N$. At mid-rapidity
$|\eta| \le 1$, however, the generated flow is quite weak and 
consistent with a zero-flow behaviour reported by the STAR and PHOBOS 
collaborations.
The $\eta$ dependence of the elliptic flow $v_2$ extracted from our 
simulation agrees well with the experimental results in the whole 
$\eta$ range for minimum bias as well as for central, mid-central and 
peripheral collisions. The transverse momentum dependence of the 
elliptic flow $v_2(p_T)$ of identified ($\pi^{\pm}, K^{\pm}, p, 
\bar{p}$) and inclusive charged hadrons has been investigated within 
the QGSM also. 
The description by the quark-gluon string model is fairly good in the 
low $p_T$ range. Here, it was shown that for identified and charged 
hadrons the $v_2$ parameter rises with increasing $p_T$ according to 
the experimental data. For higher transverse momenta 
$p_T > 1~{\rm GeV/}c$ it starts rapidly to saturate already on a 
level, which is at the largest transverse momenta roughly 50\% smaller 
than the experimentally measured $v_2$. 
On the other hand, the qualitative behaviour of the elliptic flow in 
Au+Au collisions in the overall $p_T$ range is well reproduced by the 
model. In particular, a striking feature of the QGSM is that it is 
able to describe qualitatively the different $p_T$ dependence of the 
mesonic and baryonic elliptic flow and reproduces a crossing of the 
elliptic flow for mesons and baryons at $p_T > 1.4~{\rm GeV}$ observed 
in the PHENIX experiment at RHIC. 
The centrality dependence of the integrated elliptic flow of charged 
hadrons in the QGSM agrees almost perfectly with the PHOBOS 
experimental data. This fact reflects the dominance of low $p_T$ 
hadrons.

In conclusion, the microscopic quark-gluon string cascade model 
based on the colour exchange mechanism for string formation is able 
to describe qualitatively and quantitatively well the bulk properties 
of the directed and elliptic flow measured in $\sqrt{s_{NN}} = 
200~{\rm GeV}$ Au+Au collisions at RHIC. 
The limitations of the model should not permit a perfect agreement 
for all observables. 
The only signal which is really underestimated by the QGSM is the elliptic 
flow of hadrons with high transverse momenta $p_T > 1~{\rm GeV/}c$,
although the $v_2$ of hadrons obtained within this model is 
already stronger than that of string models based on the FRITIOF 
scenario of excitation of longitudinal strings.
The most plausible explanation of this discrepancy is an anisotropic 
character of jet absorption in a hot and dense asymmetric partonic 
medium. This effect, colloquially known as {\it jet quenching\/},
is intensively studied now \cite{QM04}. Other processes related 
to the quark-hadron phase transition can be considered as well. 
The collective flow of hadrons appears to have a complex 
multi-component structure caused by rescattering and absorption 
processes in a spatially anisotropic medium. Therefore, further 
detailed investigations of the freeze-out scenario for hadrons are 
very important to understand properly the flow formation and its
evolution.

\begin{acknowledgments}
\label{acknowl}
The authors are grateful to J. Aichelin, C. Greiner, L. Csernai, 
R. Lacey and H. St\"ocker for fruitful discussions and comments.
This work was supported by the Bundesministerium f{\"u}r Bildung und 
Forschung (BMBF) under contract 06T\"U986 
and by the Norwegian Research Council (NFR).
\end{acknowledgments}




\begin{thebibliography}{00}

\bibitem{QM04}
\textit{Quark Matter 2004}, J. Phys. G \textbf{30}, S1 (2004).

\bibitem{NucBigBang}
E.V.~Shuryak, hep-ph/0011208;
R.~Stock, Nucl. Phys. A \textbf{661}, 282c (1999);
U.~Heinz, Nucl. Phys. A \textbf{685}, 414 (2001).

\bibitem{ellST01}
K.H.~Ackermann \textit{et al.}, STAR Collab., 
Phys. Rev. Lett. \textbf{86}, 402 (2001).

\bibitem{PHOB_fl}
I.~Park \textit{et al.}, PHOBOS Collab., 
Nucl. Phys. A \textbf{698}, 564c (2002);
S.~Manly \textit{et al.}, PHOBOS Collab., 
Nucl. Phys. A \textbf{715}, 614c (2003).

\bibitem{BackPHOBOS:2004}
B.B.~Back \textit{et al.}, PHOBOS Collab., 
nucl-ex/0407012.

\bibitem{Adler2003}
S.S.~Adler \textit{et al.}, PHENIX Collab., 
Phys. Rev. Lett. \textbf{91}, 182301 (2003).

\bibitem{karsch02}
F.~Karsch, Lect. Notes Phys. \textbf{583}, 209 (2002).

\bibitem{shuryak04}
E.~Shuryak, Prog. Part. Nucl. Phys. \textbf{53}, 273 (2004).

\bibitem{voloshin03}
S.A.~Voloshin, Nucl. Phys. A \textbf{715}, 379 (2003).

\bibitem{Kolb:2000fh}
P.F.~Kolb, P.~Huovinen, U.W.~Heinz, H.~Heiselberg,
Phys. Lett. B \textbf{500}, 232 (2001).

\bibitem{HKHRV2001}
P.~Huovinen, P.F.~Kolb, U.W.~Heinz, P.V.~Ruuskanen, S.~A.~Voloshin, 
Phys. Lett. B \textbf{503}, 58 (2001).

\bibitem{Ret04}
F.~Retiere, J. Phys. G \textbf{30}, S827 (2004);
T.~Hirano, {\it ibid.\/} S845.

\bibitem{Fritiof}
B.~Andersen, G.~Gustafson, B.~Nielsson-Almqvist,
Nucl. Phys. B \textbf{281}, 289 (1987).

\bibitem{dpm}
A.~Capella, U.~Sukhatme, C.I.~Tan, J.~Tran Thanh Van,
Phys. Rep. \textbf{236}, 225 (1994).

\bibitem{GRT}
V.~Gribov, Sov. Phys. JETP \textbf{26}, 414 (1968);
L.V.~Gribov, E.M.~Levin, and M.G.~Ryskin, 
Phys. Rep. \textbf{100}, 1 (1983).

\bibitem{DF_prc00}
L.V.~Bravina, A.~Faessler, C.~Fuchs, E.E.~Zabrodin,
Phys. Rev. C \textbf{61}, 064902 (2000);
Phys. Lett. B \textbf{470}, 27 (1999).

\bibitem{DF_prc01}
E.E.~Zabrodin, C.~Fuchs, L.V.~Bravina, A.~Faessler,
Phys. Rev. C \textbf{63}, 034902 (2001).

\bibitem{LPX99}
H.~Liu, S.~Panitkin, N.~Xu, 
Phys. Rev. C \textbf{59}, 348 (1999).

\bibitem{soff99}
S.~Soff, S.A.~Bass, M.~Bleicher, H.~St{\"o}cker, W.~Greiner, 
nucl-th/9903061.

\bibitem{ell_flow}
E.E.~Zabrodin, C.~Fuchs, L.V.~Bravina, A.~Faessler,
Phys. Lett. B \textbf{508}, 184 (2001).

\bibitem{agk}
V.~Abramovskii, V.~Gribov, O.~Kancheli,
Sov. J. Nucl. Phys. \textbf{18}, 308 (1974).

\bibitem{hard}
N.S.~Amelin, E.F.~Staubo, L.P.~Csernai, 
Phys. Rev. D \textbf{46}, 4873 (1992).

\bibitem{qgsm1}
A.B.~Kaidalov, Phys. Lett. B \textbf{116}, 459 (1982);
A.B.~Kaidalov, K.A.~Ter-Martirosian, 
Phys. Lett. B \textbf{117}, 247 (1982);
A.B.~Kaidalov, Surveys in High Energy Phys. \textbf{13}, 265 (1999).

\bibitem{qgsm2}
N.S.~Amelin, L.V.~Bravina, L.I.~Sarycheva, L.N.~Smirnova, 
Sov. J. Nucl. Phys. \textbf{50}, 1058 (1989);
N.S.~Amelin, L.V.~Bravina, 
Sov. J. Nucl. Phys. \textbf{51}, 133 (1990);
N.S.~Amelin, L.V.~Bravina, L.P.~Csernai, V.D.~Toneev, K.K.~Gudima, 
S.Y.~Sivoklokov, 
Phys. Rev. C \textbf{47}, 2299 (1993).

\bibitem{HuSh95}
C.M.~Hung, E.V.~Shuryak, Phys. Rev. Lett. \textbf{75}, 4003 (1995).

\bibitem{RiGy96}
D.H.~Rischke, M.~Gyulassy, Nucl. Phys. A \textbf{597}, 701 (1996).

\bibitem{Br94}
L.V.~Bravina, N.S.~Amelin, L.P.~Csernai, P.~Levai, D.~Strottman,
Nucl. Phys. A \textbf{566}, 461c (1994).

\bibitem{CsRo99}
L.P.~Csernai, D.~R{\"o}hrich, Phys. Lett. B \textbf{458}, 454 (1999).

\bibitem{Am91}
N.S.~Amelin, E.F.~Staubo, L.P.~Csernai, V.D.~Toneev, K.K.~Gudima, 
D.~Strottman, 
Phys. Rev. Lett. \textbf{67}, 1523 (1991).

\bibitem{Sorprl97}
H.~Sorge, Phys. Rev. Lett. \textbf{78}, 2309 (1997);
Phys. Rev. Lett. \textbf{82}, 2048 (1999).

\bibitem{Br95}
L.V.~Bravina, Phys. Lett. B \textbf{344}, 49 (1995).

\bibitem{VoZh96}
S.A.~Voloshin, Y.~Zhang, Z. Phys. C \textbf{70}, 665 (1996);
S.A.~Voloshin, Phys. Rev. C \textbf{55}, R1630 (1997).

\bibitem{PoVo98}
A.M.~Poskanzer, S.A.~Voloshin, Phys. Rev. C \textbf{58}, 1671 (1998).

\bibitem{Olli92+98}
J.-Y.~Ollitrault, Phys. Rev. D \textbf{46}, 229 (1992);
Phys. Rev. D \textbf{48}, 1132 (1993);
Nucl. Phys. A \textbf{638}, 195c (1998).

\bibitem{Adams:2003zg}
J.~Adams \textit{et al.}, STAR Collab.,
Phys. Rev. Lett. \textbf{92}, 062301 (2004).

\bibitem{BeltTonjes:2004jw}
M.~Belt Tonjes \textit{et al.}, PHOBOS Collab.,
J. Phys. G \textbf{30}, S1243 (2004).

\bibitem{qgsm_dir_fl_130}
L.V.~Bravina, L.P.~Csernai, A.~Faessler, C.~Fuchs, E.E.~Zabrodin, 
J. Phys. G \textbf{28}, 1977 (2002);
Phys. Lett. B \textbf{543}, 217 (2002);
L.V.~Bravina, L.P.~Csernai, A.~Faessler, C.~Fuchs, S.~Panitkin, 
N.~Xu, E.E.~Zabrodin, 
Nucl. Phys. A \textbf{715}, 665c (2003).

\bibitem{Snellings:1999bt}
R.J.M.~Snellings, H.~Sorge, S.A.~Voloshin, F.Q.~Wang, N.~Xu,
Phys. Rev. Lett. \textbf{84}, 2803 (2000).

\bibitem{Zabrodin:2004gi}
E.~Zabrodin, L.~Bravina, C.~Fuchs, A.~Faessler,
Prog. Part. Nucl. Phys. \textbf{53}, 183 (2004).

\bibitem{Wetzler:2002fi}
A.~Wetzler \textit{et al.}, NA49 Collab.,
Nucl. Phys. A \textbf{715}, 583 (2003).

\bibitem{Alt:2003ab}
C.~Alt \textit{et al.}, NA49 Collab.,
Phys. Rev. C \textbf{68}, 034903 (2003).

\bibitem{BlSt02}
M.~Bleicher, H.~St{\"o}cker, 
Phys. Lett. B \textbf{526}, 309 (2002).

\bibitem{Chen:2004vh}
L.W.~Chen, V.~Greco, C.M.~Ko, P.F.~Kolb, 
Phys. Lett. B {\bf 605}, 95 (2005).

\bibitem{MCS02}
V.K.~Magas, L.P.~Csernai, D.~Strottman, 
Nucl. Phys. A \textbf{712}, 167 (2002).

\bibitem{Back:2002wb}
B.~B.~Back {\it et al.}, PHOBOS Collab., 
Phys. Rev. Lett. {\bf 91}, 052303 (2003).

\bibitem{Back:2004je}
B.~B.~Back {\it et al.}, PHOBOS Collab., 
nucl-ex/0410022.

\bibitem{Huovinen:2003fa}
P.~Huovinen,
nucl-th/0305064.

\bibitem{Hirano:2002hv}
T.~Hirano, K.~Tsuda,
Nucl. Phys. A {\bf 715}, 821 (2003).

\bibitem{Hirano:2002ds}
T.~Hirano, K.~Tsuda,
Phys. Rev. C {\bf 66}, 054905 (2002).

\bibitem{Hirano:2001eu}
T.~Hirano,
Phys. Rev. C {\bf 65}, 011901 (2002).

\bibitem{Bravina:2004td}
L.~Bravina, K.~Tywoniuk, E.~Zabrodin, G.~Burau, J.~Bleibel, C.~Fuchs, 
A.~Faessler,
hep-ph/0412343.

\bibitem{Adams:2004bi}
J.~Adams \textit{et al.}, STAR Collab., 
nucl-ex/0409033.

\bibitem{AdamsSTAR:2004}
J.~Adams \textit{et al.}, STAR Collab., 
Phys. Rev. Lett. \textbf{92}, 112301 (2004).

\bibitem{MV03}
D.~Molnar, M.~Gyulassy, Phys. Rev. C \textbf{62}, 054907 (2002);
D.~Molnar, S.A.~Voloshin, Phys. Rev. Lett. \textbf{91}, 092301 (2003).

\bibitem{GKL2003}
V.~Greco, C.M.~Ko, P.~Levai, Phys. Rev. C \textbf{68}, 034904 (2003).

\bibitem{FMNB2003}
R.J.~Fries, B.~M{\"u}ller, C.~Nonaka, S.A.~Bass, 
Phys. Rev. C \textbf{68}, 044902 (2003).

\bibitem{HY2003}
R.C.~Hwa, C.B.~Yang, Phys. Rev. C \textbf{67}, 034902 (2003).

\bibitem{LoSn01}
I.P.~Lokhtin, S.V.~Petrushanko, L.I.~Sarycheva, A.M.~Snigirev,
Pramana \textbf{60}, 1045 (2002); 
I.P.~Lokhtin, L.I.~Sarycheva, A.M.~Snigirev,
Phys. Lett. B \textbf{537}, 261 (2002).

\bibitem{GBA04}
F.~Gastineau, E.~Blanquier, J.~Aichelin, 
hep-ph/0404207.

\bibitem{Snellings:1999gq}
R.J.M.~Snellings, A.M.~Poskanzer, S.A.~Voloshin, 
nucl-ex/9904003.

\end{thebibliography}
\end{document}